\documentclass[aps,preprint]{revtex4}
\usepackage{epsfig}
\begin{document}

\title{\bf Comparison between ab-initio and phenomenological modeling of the exchange couplings in diluted magnetic semiconductors: the case of  $Zn_{1-x}Cr_{x}Te$.
\footnote{To appear in ``Proceedings of the International Conference MSM05, Agadir Maroc September 2005'', Physica Status Solidi}
}

\author{Georges ~Bouzerar$^{1}$, Richard Bouzerar$^{2}$, Josef ~Kudrnovsk\'y'$^{3,4}$
and Timothy Ziman$^{4}$}

\address{
$^{1}$ Laboratoire Louis N\'eel  25 avenue des Martyrs BP 166 38042 Grenoble Cedex 09
France.\\
$^{2}$Laboratoire de Physique de la Mati\`ere Condens\'ee Universit\'e de Picardie Jules Verne,33 rue Saint Leu 80039 Amiens Cedex 01 France\\
$^{3}$ Institute of Physics  Academy of Sciences, Na Solvance 2 CZ-182 21 Prague 8 Czech Republic \\
$^{4}$Institut Laue Langevin
BP 156 38042 Grenoble
France.\\
}

\begin{abstract}
Using a recently developed semi-analytical method (Self-Consistent Local RPA or SC-LRPA) we study the stability of the ferromagnetic phase in diluted magnetic systems where the exchange coupling between magnetic impurities are of  RKKY form. A short discussion of the relevance of these calculations with respect to the ferromagnetism observed in diluted ferromagnetic materials is provided. Then, within a two step approach, we study ferromagnetism in $Zn_{1-x}Cr_{x}Te$. In the first step of our study, we calculate the magnetic couplings between Mn impurities within the LDA. In the second step, we diagonalize the resulting effective Heisenberg Hamiltonian using the SC-LRPA.
We also compare, when available, our calculations with Monte Carlo simulations and experimental measurements.
\end{abstract} 

\maketitle

%]
Diluted Magnetic Semiconductors materials have attracted considerable attention because of their technological potential in the domain of spintronics. These materials are particularly promising since a small concentration of magnetic impurities can already give rise to a relatively high Curie temperature \cite{Ohno,Edmonds}. The key issue to obtain quantitative and reliable values of the Curie temperature is twofold. First, one has to be able to provide correct values for the exchange couplings between magnetic impurities. Secondly, one must treat   properly the resulting effective Heisenberg Hamiltonian which describes the interaction between randomly distributed magnetic impurities on the lattice. By properly, we mean provide an accurate treatment of (i) thermal fluctuations 
and (ii)  effects of disorder in the position of the impurities. 
Different approaches have been developed for the determination of the exchange integrals in III-V materials.
There are essentially two different kind of methods. First, simple phenomenological approaches \cite{Dietl,Abolfath} in which the band structure of the host material is described in a realistic manner by a Kohn-Luttinger Hamiltonian for the valence band states and a coupling $J_{pd}$ (or $V_{pd}$) between the itinerant carrier and the localized magnetic impurities.
In these approaches, the coupling between localized magnetic impurity and the itinerant carrier is only treated {\it perturbatively}. However, within a simple single band model in which the local $J_{pd}$ coupling was treating non perturbatively within CPA, it has been shown that the perturbative treatment which leads to $J_{ij} \propto J_{pd}^{2}$ and thus to $T_{C}\propto J_{pd}^{2}$ is only restricted to very weak coupling\cite{Georges-cpa}. 
A second class of  approaches are based on first principle calculations.
In these approaches, the coupling between d states of the magnetic impurities and host states is treated non-perturbatively,  and this leads to realistic values  of the exchange couplings. The main, and great, advantage of these approaches is that the couplings do not depend on adjustable parameters.

This manuscript is organized as follows. The first part is devoted to the study of the effects of thermal fluctuations and disorder on the stability of the ferromagnetic region as the density of itinerant carriers in 
diluted magnetic systems, with assumed RKKY-like  exchange couplings. In view of our results, we discuss the relevance of such a
phenomenological model to simulate the exchange couplings in actual diluted ferromagnetic semiconductors.
 In the second part, using realistic first principle calculations, we calculate the exchange couplings within the Tight-Binding LMTO approach in Cr doped ZnTe semiconductors. We then calculate the Curie temperature as a function of the magnetic impurity concentration for this material.

Throughout this paper, we present calculations of the Curie temperature  obtained using a recently developed Self-Consistent Local Random Phase Approximation (SC-LRPA) \cite{epl2005,prb2005}. The  Curie temperatures calculated within SC-LRPA appear to be in very good agreement with Monte Carlo simulations where the same exchange couplings have been used \cite{MonteCarlo}. However the SC-LRPA has several important advantages with respect to Monte Carlo calculations: it allows (i) to derive a direct semi-analytical expression for the Curie temperature (no extrapolation is required), (ii) the spins can be either quantum or classical, (iii) the calculations are very fast and (iv) correlation in the disorder can be included without any complication
\cite{apl2004} . Within SC-LRPA the self-consistency is usually achieved after approximately 10-15 iterations, i.e. it requires the same number of diagonalizations of a $N_{imp}\times N_{imp}$ complex matrix. For comparison, for a given configuration of disorder, our calculations of Tc are at least 3 orders of magnitude faster than Monte Carlo calculations. Of course speed is not
the only point; 
in the low temperature phase the SC-LRPA  includes quantum effects that cannot
be  simulated classically. We now summarize the main steps of the SC-LRPA treatment. The disordered Heisenberg Hamiltonian which we consider is

\begin{eqnarray}
H_{eff}=-\sum_{ij} J_{ij} {\bf S}_{i}\cdot {\bf S}_{j}
\end{eqnarray}
The sum runs only over pairs of sites occupied by magnetic impurities.
The spins are quantum with value S. In the case of classical spins, we will properly
perform the limit $S \rightarrow \infty$ at the end of the calculations.
We now define the following retarded Green's function,
\begin{eqnarray}
G_{ij}(\omega)=\int_{-\infty}^{+\infty} G_{ij}(t) e^{i\omega t} dt
\end{eqnarray}
where $G_{ij}(t)=-i\theta(t)\langle [S^{+}_{i}(t);S^{-}_{j}(0)] \rangle$.

After standard Tyablicov decoupling of the equation of motion of  $G_{ij}(\omega)$ we obtain,

\begin{eqnarray}
(\omega-h_{i}^{eff})G_{ij}(\omega)=2 \langle S_{i}^{z} \rangle \delta_{ij} -
 \langle S_{i}^{z} \rangle \sum_{l} J_{il}G_{lj}(\omega)
\end{eqnarray} 

where the local effective field is,

\begin{eqnarray}
h_{i}^{eff}= \sum_{l} J_{il} \langle S_{l}^{z} \rangle 
\end{eqnarray}

For a given temperature and fixed disorder configuration, the local magnetization $\langle S_{i}^{z} \rangle$ has to be determined self-consistently at each impurity site. In order to close the set of equation we use Callen expression, which relate the local Green's function at site i to the local magnetization at this site \cite{Callen}.
\begin{eqnarray}
 \langle S_{i}^{z} \rangle = \frac{(S-\Phi_{i})(1+\Phi_{i})^{2S+1}+(S+1+\Phi_{i})\Phi_{i}^{2S+1}}
{(1+\Phi_{i})^{2S+1}-\Phi_{i}^{2S+1}}
\end{eqnarray}
The local effective magnon occupation number reads,

\begin{eqnarray}
\Phi_{i}=\frac{-1}{2\pi \langle S_{i}^{z} \rangle} \int_{-\infty}^{+\infty} \frac{Im G_{ii}
(\omega)}{exp(\omega/kT)-1}d\omega
\end{eqnarray}

The previous set of equations allows us now to determine, at each temperature, the local magnetization at each impurity site, and the dynamical properties such as the dynamical structure factor.

To determine the value of the Curie temperature we take the limit $ \langle S_{i}^{z} \rangle \rightarrow 0$ in the previous set of equations. We obtain,

\begin{eqnarray}
k_{B}T_{C}=\frac{1}{3} S(S+1) \frac{1}{N_{imp}} \sum_{i} \frac{1}{F_{i}}
\end{eqnarray}
where,

\begin{eqnarray}
F_{i}=\int_{-\infty}^{+\infty} \frac{A_{ii}(E)}{E} dE
\end{eqnarray}
The local spectral function $A_{ii}(E)=-\frac{1}{2\pi} Im(\frac{G_{ii}(E)}{{\lambda_{i}}})$. In the previous equation the factor $\lambda_{i}=lim_{T \rightarrow T_{C}}  \frac{\langle S_{i}^{z} \rangle}{m}$ (m is the magnetization averaged over all impurity sites). It is interesting to note that the Curie temperature can be re-expressed in term of the eigenfunctions and eigenvalues of the following effective Hamiltonian, where the matrix elements are,

\begin{eqnarray}
(H_{eff})_{ij}=-\lambda_{i} J_{ij} + \delta_{ij} \sum_{l} \lambda_{l}J_{il}
\end{eqnarray}

This lead to,
\begin{eqnarray}
F_{i}=\sum_{\alpha}  \frac{|\langle i |\Psi_\alpha \rangle|^2}{ E_{\alpha}}
\end{eqnarray}

Thus the Curie temperature becomes,

\begin{eqnarray}
T^{sc}_{C}=\frac{1}{3 N_{imp}} S(S+1) \sum_{i} (\sum_{\alpha} \frac{|\langle i |\Psi_\alpha \rangle|^2}{ E_{\alpha}} )^{-1} 
\end{eqnarray}

This expression implies that the nature (extended/localized) of the eigenstate will have  important effects on the magnetic excitation spectrum and on the Curie temperature. A detailed discussion of these aspects will be provided elsewhere \cite{inpreparation}.

%{\bf Partie RKKY-partie figure Richard.}
We now turn to the commonly used  model of dilute magnetic systems where we 
place spins randomly, and independently,
on a host lattice, and  assume an  interaction between each pair of 
spins of RKKY form with some density
of carriers mediating the exchange. The carrier density is an important parameter, as it determines
the scale of oscillation of the exchange.
In Fig.1, we have plotted the Curie temperature for a fixed concentration of 
magnetic impurities  as a function of the carrier density, normalized
with respect to the concentration of  impurities. We assume that the exchange couplings are of RKKY form and we including,
the effects of disorder in an averaged  way, i.e.  by including an exponential
damping of the couplings. 
Thus the interaction is taken to be 
\begin{eqnarray}
J_{\rm ij} &=&  J_{0} \exp({-{\frac r  R_{0}}}) \frac{({\sin (2k_{F}r)-2k_{F}r\cos (2k_{F}r)})} {(r/a)^{4}} 
\end{eqnarray}
with Fermi vector $k_F$ determined by the density
$k_{\rm F} = (3 \pi ^2 n_{c})^{\frac 1 3}$ where $n_{c}=\gamma x$ is the density of carriers for impurity concentration $x$.

\begin{figure}[tbt]
\includegraphics[width=.60\textwidth,angle=-90]{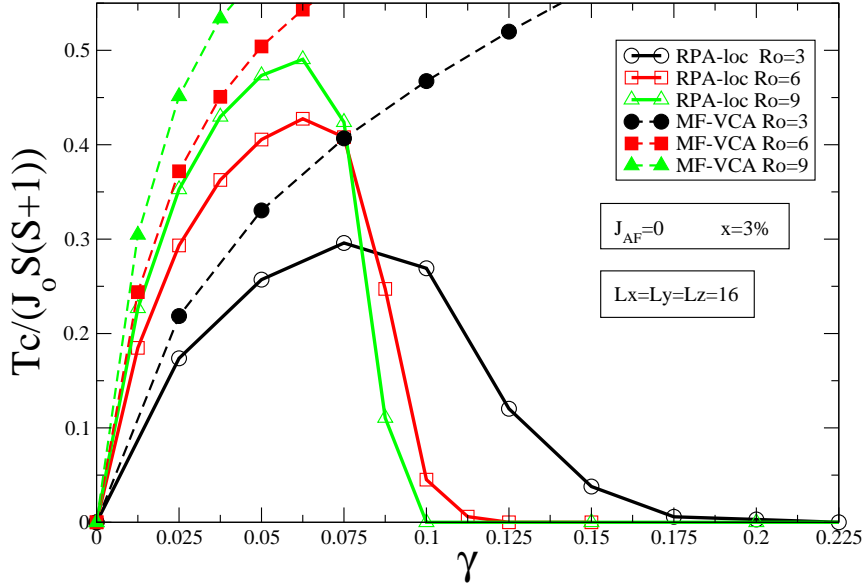}
\caption{Curie temperature as a function of the carrier density for a fixed concentration of magnetic impurities (3\% ) and exchange couplings of RKKY form (see equation 12). The exponential cut-off in interactions
is defined by the screening parameter $R_0$   
.} 
\label{fig.1}
\end{figure}

The curves of Curie temperature versus carrier
concentration are shown for three different values of damping. Note that we do not consider in this work the effect
 of a short range super-exchange contribution; this is discussed in detail in 
Ref.\cite{Richardetal-rkky}.  
For each case we show, for comparison (dotted lines), the Curie temperature
estimated from mean-field Virtual Crystal Approximation (MF-VCA).
This approximation has often been used, but, unlike ours, replaces
the random lattice of impurities by a translationally invariant
effective medium.
We observe that as the carrier concentration increases from zero, 
our calculated  Curie Temperatures have  very different behaviour from the mean-field theory. Most dramatically, rather than increasing monotonically
 with carrier density, the Curie temperature 
saturates and vanishes at a density of carriers much smaller than the density of impurity moments. This is significant in that in doped Ga(Mn)As, for example,
it is clearly seen from simultaneous transport and magnetic measurements
that the highest Curie temperatures are obtained as the samples are annealed,
giving carrier densities per moment approaching unity. Thus we conclude
that simple RKKY interactions are {\it not} appropriate to describe
the ferromagnetism in doped III-V semiconductors. The reasons for 
this had been obscured by the use of mean-field theories. The reason
that the SC-LRPA shows the instability at higher carrier concentration is that
includes the effect of the oscillating tail of the exchange
constants, i.e. the frustration in the magnetic interactions. As damping becomes more short range, i.e. $R_0$ becomes smaller, the frustrating tail of interactions is reduced, as seen in Figure 1, but this is not enough to suppress the instability.

\par

\begin{figure}[hbt]
\includegraphics[width=.53\textwidth]{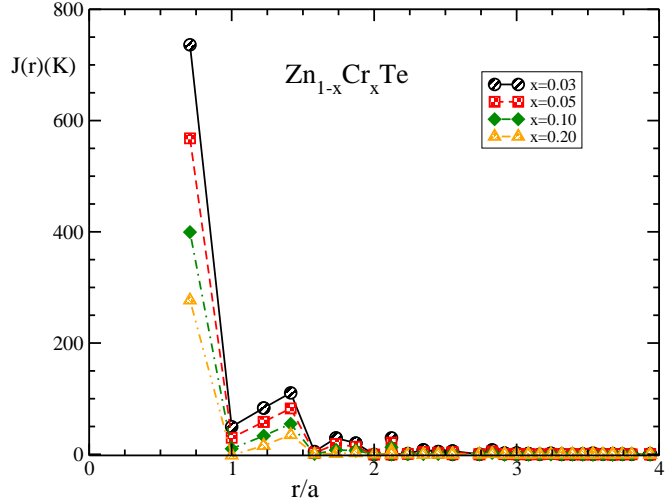}
\caption{Calculated Heisenberg exchange couplings (in Kelvin) as a function of the distance (in units of the lattice constant) between magnetic impurities for different concentration of Cr  $x=3 \%$ , 5 $\%$, 10 $\%$ and 20 $\%$ .} 
\label{fig.2}
\end{figure}

We learn from this that  in order to explain observed ferromagnetism in diluted magnetic semiconductors we need to model the magnetic interactions more accurately. An approach
that has proved successful, both in doped III-V\cite{epl2005} and II-VI
materials, as we shall see below, is to use {\it ab initio} methods to estimate
the magnetic couplings at all distances including effects of the (disordered) band structure,
correlations, screening treated within, for example Local Density Approximation (LDA). As the magnetic couplings can be calculated 
by a magnetic force theorem, i.e. in principle from the energies of the ground
state perturbed by weak magnetic fields, they should be
accurate. With the resultant Hamiltonian, which is specific to 
both the host material and the doping concentration (since disorder
and doping effects
are taken into account in the band structure), we will apply
our method of SC-LRPA to estimate the Curie temperature.
Within the class of II-VI semiconductors we  turn now to the experimentally promising candidate for room temperature ZnTe doped with Cr.
 
 First we show the exchange integrals between Cr moments deduced from our
{\it ab initio} calculations. In Fig.2, we have plotted the magnetic exchange
 integrals between Cr impurities in ZnTe as a function of the distance, for 
different concentration. First, we clearly see that the frustration,
which, we argued, could strongly suppress ferromagnetism in an RKKY model,
is absent here. This is related to the fact that the Chromium dopants
are close to the configuration Cr$^{2+}$, i.e. do not, to a first approximation,
introduce carriers into the II-VI host. In fact the band structure shows ``half-metallicity'':
a finite density of states at the Fermi energy for the majority carriers
and a gap for the minority carriers\cite{Wang}. 
The couplings, while oscillating
in magnitude, remain essentially ferromagnetic and
exponentially damped. The ferromagnetic nature at short distances
is attributed to the strong
hybridization of the Cr 3d levels with the Te 5 p states \cite{Wang}. The damping 
has different origins, the effects of disorder (treated within CPA), 
the half-metallicity and the effects of the d levels.
As the concentration of Cr increases, the couplings decrease at short distances
keeping the same position of local minima.\\

\begin{figure}[hbt]
\includegraphics[width=.53\textwidth]{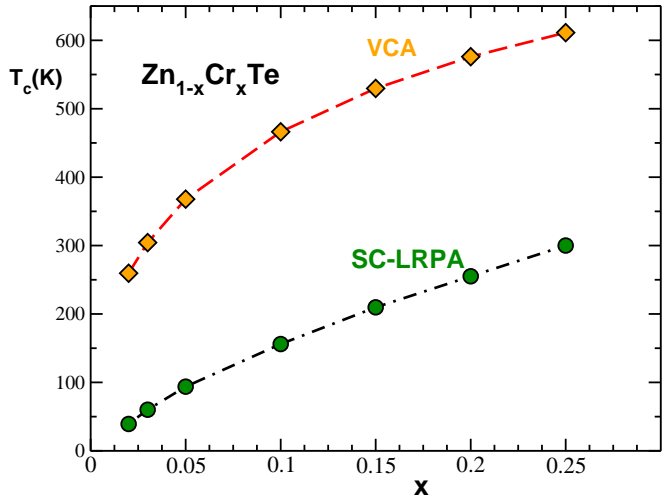}
\caption{Calculated Curie temperature within Self-Consistent Local RPA and Mean Field VCA for $Zn_{1-x}Cr_{x}Te$ as a function of $x$. } 
\label{fig.3}
\end{figure}

In Fig.3 we show the variation of the Curie temperature calculated within SC-LRPA as a function of the magnetic impurity concentration.
A systematic average over at least 100 configurations of disorder has been 
taken for each concentration.  For comparison,
we also show the Mean-Field VCA results. While qualitatively the differences
are not as dramatic as in Fig.1 in that both mean-field and our
curve are monotonically increasing, {\it quantitatively} the predictions
are very different. Most importantly, we believe our theory should
be a reliable guide to the highest Curie temperatures that should
be obtained for each concentration\cite{epl2005}.,
In particular we predict
that room temperature should require concentrations around 25 \%; the mean field
VCA
 gives much lower values for this minimum concentration, about 2\%. 
Experimentally Saito et al\cite{Saito}
have found T$_C = 300\  {\frac + -}\  10$ for thin films with $x \approx 0.2$ which 
is close to  our prediction(T$_C(x=0.2) = 265$). 
Monte Carlo methods\cite{MonteCarlo} for Zn(Cr)Te using the same exchange
couplings were made for  a few  concentrations and  are close
to our curve. For example the Monte Carlo result is  (T$_C(x=0.2) \approx 300$).  
We would
conclude that the samples measured are close to the optimal Curie temperature
for the concentration of  uncorrelated impurities.

In conclusion, we have shown that the region of stability of ferromagnetism in diluted systems where the exchange
couplings are RKKY-like is very narrow even when there is  large damping.
These couplings are very different from those obtained in ferromagnetic 
diluted semiconductors by means of first-principle calculations, and 
thus appear to be inappropriate to explain the wide region  (in  carrier concentration) 
of stability
usually observed in these materials. We have calculated the variation of the Curie temperature in $Zn_{1-x}Cr_{x}Te$ 
as a function of Cr density and found that room temperature can be achieved for densities above $x \approx 0.25$.
This is important for this particular material and 
also illustrates how, by coupling first principle calculations of the exchange couplings with SC-LRPA, we may make useful predictions for  a range of
important new materials.

\end{document}